\documentclass[aps,nofootinbib,prd,epsf]{revtex4}
\usepackage[latin1]{inputenc}
\usepackage{amssymb}
\usepackage{float}
\usepackage{graphicx}
\usepackage{amsmath,amssymb,amsbsy,bm}

\usepackage{subfigure}

\usepackage{graphicx}
\usepackage{epsfig}

\newcommand{\beq}{\begin{equation}}
\newcommand{\eeq}{\end{equation}}
\newcommand{\bea}{\begin{eqnarray}}
\newcommand{\eea}{\end{eqnarray}}
\def\bi{\begin{itemize}}
\def\ei{\end{itemize}}

\newcommand{\imag}{\mathrm{i}}

\newcommand{\diracd}{\delta}
\renewcommand{\d}{\mathrm{d}}
\newcommand{\e}[1]{\mathrm{e}^{{#1}}}
\newcommand{\vect}[1]{\bm{\mathrm{{#1}}}}

\newcommand{\even}[1]{{\mbox{\scriptsize ${#1}$ even}}}

\newcommand{\Prob}{\bm{\mathbb{P}}}

\newcommand{\Sigmar}{\Sigma_{\mathcal{R}}}
\newcommand{\G}{\mathbb{G}}

\newcommand{\R}{\mathcal{R}}
\newcommand{\Rsp}{\mathsf{R}}

\newcommand{\window}{\mathcal{W}}
\newcommand{\kmax}{k_{H}}

\newcommand{\ps}{\mathcal{P}}

\newcommand{\bx}{{\bf{x}}}
\newcommand{\by}{{\bf{y}}}
\newcommand{\br}{{\bf{r}}}
\newcommand{\bk}{{\bf{k}}}
\newcommand{\hr}{{\hat{r}}}

\newcommand{\thetak}{\theta_k}

\newcommand{\phik}{\phi_k}
\newcommand{\K}{\mathsf{K}}

\begin{document}

\title{Probability of primordial black hole formation and its dependence on
 the radial profile of initial configurations}

\author{J.~C. Hidalgo$^{\dag}$ and A.~G. Polnarev $^{\ddag}$}
\email[\dag]{c.hidalgo@qmul.ac.uk}
\email[\ddag]{a.g.polnarev@qmul.ac.uk}
\affiliation{Astronomy Unit, School of Mathematical Sciences,\\
  Queen Mary College,  University of London, Mile End  Road,\\
  London E1 4NS, United Kingdom }

\date{\today}

\begin{abstract}

In this paper we derive the probability of the radial profiles of
spherically symmetric inhomogeneities in order to provide
an improved estimation of the number density of primordial black holes
(PBHs). We demonstrate that the probability of PBH formation depends
sensitively on the radial profile of the initial configuration.
We do this by characterising this profile with two parameters chosen
heuristically: the amplitude of the inhomogeneity and the second
radial derivative, both evaluated at the centre of the
configuration. We calculate the joint probability of initial
cosmological inhomogeneities as a function of these two parameters and
then find a correspondence between these parameters and those used in 
numerical computations of PBH formation. Finally, we extend our
heuristic study to evaluate the probability of PBH formation taking
into account for the first time the radial profile of curvature
inhomogeneities.   

\end{abstract}


\maketitle
\section{Introduction}

The idea that large amplitude matter overdensities in the universe could have
collapsed through self-gravity to form  primordial
black holes (PBHs) was first put forward by Zeldovich and Novikov
\cite{zeldovich66}, and then independently by Hawking \cite{hawking71},
more than three decades ago. This theory suggests that large amplitude
inhomogeneities in the very early universe overcome  
internal pressure forces and collapse to form black holes.
A lower threshold for the amplitude of such inhomogeneities
$\delta_{\textrm{th}} \equiv (\delta \rho / \rho)_{\textrm{th}}$, was
first provided by Carr \cite{carr75}, giving $ \delta_{\textrm{th}}
\approx 1/3 $  at the time of radiation domination. This value was
found by  comparing the Jeans length of the overdensity with the scale
of the cosmological horizon at the time of formation.  

The mass fraction of the universe turning into PBHs of mass $M$ at
their formation time, $\beta_{\textrm{PBH}}(M)$, is computed 
using the probability density function (PDF) of the relevant field of
inhomogeneities, which is provided by the cosmological theory. The mass
fraction $\beta_{\textrm{PBH}}(M)$ is customarily given by the
integral of this PDF over the amplitude $\delta \equiv \delta\rho /
\rho$, with a lower limit equal to $\delta_{\textrm{th}}$
\cite{press74,carr75}.   

The probability of PBH formation is a useful tool to constrain
the mean amplitude of inhomogeneities on scales which cannot be probed
by any other methods. The PBH contribution to the energy density
increases with time during the radiation-dominated epoch. For this
reason, the  PBHs formed considerably
before the end of radiation domination are the most relevant to cosmology
\cite{lidsey94,liddle98,sendouda06,zaballa06,bugaev06,kohri07}. We
will focus our study on these kind of PBHs and assume that the
background matter at the time of PBH formation is
radiation-dominated. To make this cosmological tool more precise, we must
improve the calculations of the probability of PBH formation. This
demands a more accurate evaluation of the threshold value
$\delta_{\textrm{th}}$, or the equivalent curvature inhomogeneity
$\R_{\textrm{th}}$
\cite{bicknell79,niemeyer97,jedamzik99,shibata99,hawke02,musco04}. 
In search of these values, it was evident that the
process of PBH formation depends on the pressure gradients in the
collapsing configuration in addition to the amplitude
\cite{nadezhin78,shibata99,musco07}. It was also found that such pressure
gradients can modify the value of $\delta_{\textrm{th}}$
significantly. Hence, when calculating the probability of PBH
formation, one should consider the shape and radial profile of the
initial configurations. These profiles are directly related with the
internal pressure gradients. This is the main motivation for the
present work. Previous studies, concerning 
the probability of PBH formation, take the amplitude of
perturbations to be the only parameter determining the probability
density. In addition to that, we include for the first time a
parameter related to the slope of curvature profile at the edge of the
configuration 
\footnotemark\footnotetext{In the context of dark matter haloes the
  question of initial profiles is effectively irrelevant because
  galaxies are formed from pressureless configurations. The density profiles
  and shapes of virialized haloes result from the evolution of the
  initial high peaks and are not linked to the profile of initial
  configurations that we investigate here (see e.g. \cite{cooray02}
  for a review on the profiles of dark matter haloes and
  \cite{maccio08} for some recent results on this topic).}.  

In this paper, we calculate the probability of finding a curvature
configuration with a given radial profile. As follows from
\cite{polnarev81,polnarev82, khlopov83,polnarev85,zabotin87}, PBH
formation takes place only from nearly spherical configurations, so in
this paper we restrict ourselves to the spherically symmetric case.
In this first approximation we describe the radial profiles by
introducing two parameters: the central amplitude of the curvature
inhomogeneity $\R(\br = 0)$ and the second radial derivative at
the centre $\R''(\br = 0)$, which is chosen mainly to avoid
technical difficulties. The introduction of these parameters is a
first step towards the full parametrisation of profiles in terms of
even derivatives at the centre of configurations, i.e. in terms of
$\R^{(2n)}(0)$ (the odd derivatives $\R^{(2n + 1)}(0)$   are all zero
due to the assumed spherical symmetry). In the future, with the
results from more accurate codes simulating the formation of PBHs, we
will have at hand a larger number of conditions for the collapse of a
curvature profile. An equal number of parameters will be required for
the complete description of these profiles and the probability of
finding them. In the meantime, only families of curvature profiles
described by two parameters are available. We consequently limit
ourselves to the two-parametric description of initial curvature
profiles.

The central amplitude $\R(0)$ has been used in previous calculations of
gravitational collapse \cite{shibata99}, and the probability of PBH
formation \cite{green04,zaballa06}. In the present paper we compute the
probability to find a given configuration as a function of the two
parameters $[\R(0),\R''(0)]$. We subsequently illustrate how this
two-parametric probability is used to correct the probability of PBH 
formation. Such an exercise is presented for illustration
purposes. The results, based on a non-rigorous but physically
meaningful determination of the parameters which describe the initial
profiles, show how the corrections to $\beta_{\mathrm{PBH}}$ are
significant and they will be considered in more detail in future
studies of PBH formation. 

This paper is organised as follows. In Section
\ref{prob:construction} we calculate the joint probability
distribution for $\R(0)$ and $\R''(0)$. In Section \ref{metrics}, we
relate these parameters to those used in the most recent numerical
computations of PBH formation. In Section
\ref{prob:comparison} we present the total probability of PBH
formation, integrating the probability distribution derived in Section
\ref{prob:construction} over the relevant region of
parameter space $\left[\R(0)\,,\,\R''(0)\right]$. In Section
\ref{conclusion} we summarise our results and discuss
future research in this area. 

\section{Probability of radial profile parameters of cosmological
  perturbations}\label{prob:construction}

The most striking prediction of the theory of cosmological inflation
is that initial quantum fluctuations are transfered into the
inhomogeneities and structures observed in the universe today. After
inflation, the universe is mostly flat with inhomogeneities of small
amplitude on average (for a review, see \cite{liddle00}). Some
of the inhomogeneous regions, present high amplitude (non-linear
inhomogeneities) and these are the object of study in the present paper.
Formally, the high amplitude inhomogeneous profiles describing
configurations which collapse into PBHs are not perturbations.
However, such regions are included in the statistics of random
primordial curvature perturbations. That is, the statistics of random
fields can be used to estimate the probability of finding high amplitude
inhomogeneities.

To describe large-amplitude inhomogeneities, we consider the non-linear
curvature field $\R(t,r)$, as first described in \cite{salopek90}.
The non-linear curvature $\R(r,t)$ represents the relative expansion
of a given local patch of the universe with respect to its neighbouring
patches \cite{lyth05}. It is described by the metric
\begin{align}
  \d s^2 = - N^2(t,\vect{r}) \, \d t^2 + a^2(t) \e{2\R(t,\vect{r})}
  \tilde{\gamma}_{ij} 
  (\d r^i + N^i(t,\vect{r}) \, \d t)( \d r^j + N^j(t,\vect{r}) \, \d
   t), \label{0.1}
\end{align}

\noindent where $a(t)$ is the scale factor and the gauge dependent
functions $N$ and $N^i$ are the lapse function and shift vector,
respectively. These variables are determined by algebraic
constraint equations in terms of the matter density$~\rho$ and pressure$~p$ and
the metric variables $\R,~a$ and $\tilde{\gamma}_{ij}$.

In this work we consider the non-linear configurations which
correspond to non-zero and large $\R$ inside some restricted volume,
and zero outside, where the expansion of the universe follows the
background Friedmann-Robertson-Walker (FRW) solution.
There are several advantages of working with metric
\eqref{0.1}. First, $\R$ is defined as a gauge-invariant combination of metric
and matter variables \cite{wands00}. Second, with the aid of the
gradient expansion of the metric quantities
\cite{salopek90,deruelle94,rigopoulos04,langlois05},
$\R(\br,t)$ is presented in the Einstein equations in a non-perturbative
way. Finally, $\R$ does not depend on time for  scales larger than the
cosmological horizon, as proved in \cite{lyth05,langlois05b} and in a more
general case in \cite{lifshitz63}. In the
present paper we work in the superhorizon r\'egime, where the field
$\R(\br)$ can be assumed to be time-independent.

The primordial field of random perturbations we use follows a Gaussian
probability distribution. In other words, we assume that the
probability of finding a curvature perturbation $\R$ of mean amplitude
$\vartheta$ is given by the probability density function (PDF) 
 \begin{align}
   \label{gaussian:prob}
    \Prob[\vartheta] \propto
    \exp{\left(-\frac{\vartheta^2}{2\Sigmar^2}\right)},
\end{align}

\noindent where $\Sigmar$ is the dispersion of the perturbation field
$\R$, defined in terms of the two-point correlation function by
\begin{align}
  \label{intro:variance}
  \Sigmar^2(r_H) = \langle \R(r)\R(r) \rangle = \int \d \ln k
  \;\window^2(k,k_H) \ps(k), 
\end{align}

\noindent  where $\window(k,k_H)$ is the window function which
smooths the field over spherical regions of size $r_H = 2\pi /
k_H$, the Hubble radius. The power spectrum  of $\R$, $\ps(k)$, is an
output of the underlying cosmological model, as reviewed in
\cite{liddle00}.  

In more general cases, PDFs include the contribution of higher-order
correlations (i.e. $\langle\R\R\R\rangle$ and all other correlations). To
calculate such PDFs, a new formalism is required, such as that
developed in \cite{sh06}. Several studies have shown that the
non-Gaussian correlations can sensibly modify the PDF of the amplitude
of perturbations and consequently the  number count of astrophysical
objects \cite{loverde07,silk07,matarrese00} and PBHs
\cite{bullock97,ivanov98,hidalgo07,saito08} when large
non-Gaussianities arise in the primordial field of curvature
fluctuations \cite{seery05,lyth05b,chen06a,chen06b}. Here however, we
restrict ourselves to the Gaussian case where the PDF presents the
form of Eq.~\eqref{gaussian:prob}.

In the following we calculate the joint probability of finding an
amplitude $\R(0)$ and the second derivative 
\begin{align}
  \R''(0) = \left[\frac{\partial^2}{\partial r^2}\R(r)\right]_{r = 0},
\end{align}

\noindent at the centre, using the method developed in
\cite{sh06}. In order to compute the probability of a specific
property of $\R(\br)$, we integrate the original PDF, which
encodes all the information about the field, with the Dirac
$\delta$-functions of relevant arguments. In particular, the
probability that  $\R(0) = \vartheta_0$ is given by,
\begin{equation}
  \mathbb{P}(\vartheta_0) = \int[d\R]
  \mathbb{P}(\R) \,\delta\left[\R(0) - \vartheta_0
    \right], \label{delta:integral}
\end{equation}

\noindent were $[d\R]$ indicates integration over all possible
configurations $\R(\bk)$ in Fourier space. Hereafter we consider $\R(0)$ and
$\R''(0)$ as statistically independent parameters. Hence, the
probability of having $ \R''( 0)= \vartheta_2$, is given by the integral
\begin{equation}
  \mathbb{P}(\vartheta_2) = \int[d\R]
  \mathbb{P}(\R)\, \delta\left[\R''(0) - \vartheta_2
  \right]. \label{delta:integralddr}
\end{equation}

\noindent In the rest of this section we show roughly how this method
works. The details of the following results are presented in
appendices \ref{appA} and \ref{appB}. First we expand the smoothed
curvature perturbation profile $\R(\br)$ in terms of spherical
harmonic functions: 
\begin{align}
      \R({\br}) = &{\displaystyle {\int}
    }\frac{d^3k}{(2\pi)^3}\R({\bf k})\exp{(\textrm{i}{\bf k\cdot r})},
    \label{1.3a}
\end{align}

 \noindent with
\begin{align}
  \R(\bk) = &\sum_{\ell = 0}^\infty\,
    \sum_{m = - \ell}^{\ell}\,  \sum_{n = 1}^\infty\,
    \R^m_{\ell|n}\, Y_{\ell m}(\theta,\phi)
    \psi_n(k). \label{1.3b}
\end{align}

\noindent Here $Y_{\ell m}$ are the usual spherical harmonics on the
unit 2-sphere and $\psi_n(k)$ are a complete and orthogonal set of
functions in an arbitrary finite interval $0\,<\, k \,<
\,\Lambda$ (for an explicit expression of $\psi(k)$ see appendix
\ref{appA}). It is worth mentioning that the cutoff $\Lambda$ is
imposed to artificially compactify the momentum space. This allows us
to provide an explicit definition of the functions $\psi_n(k)$ and a
complete set of functions $\psi$ for the expansion of $\R(\bk)$. In
turn this condition allows a regularisation of the path integral
$\int[d\R]$  by considering the harmonic expansion \eqref{1.3b} in a
finite interval in Fourier-space $0 < k < \Lambda$. At the end of the
calculation we can take the limit $\Lambda \to \infty$ and the results
will remain unchanged. The coefficients in the expansion are
generically complex, so we separate real and  imaginary part
introducing $\R^m_{\ell|n} = a^m_{\ell|n} +
\textrm{i}b^m_{\ell|n}$. The reality condition for the curvature
field,  $\bar{\R}(\bk) = \R(-\bk) $, is met when 
\begin{align}
  \label{1.4}
    a^{-m}_{\ell|n} = (-1)^{\ell+m}a^m_{\ell|n}, \\
    b^{-m}_{\ell|n} = (-1)^{\ell+m+1}b^m_{\ell|n}.\label{1.41}
\end{align}

\noindent In particular, the $m = 0$
modes require $a^{0}_{\ell|n}$ and $b^{0}_{\ell|n}$  to be
zero for odd and even $\ell$, respectively. After evaluating the expansion
\eqref{1.3a}-\eqref{1.3b} at $\R(\br = 0)$, we can use the relation,
\begin{align}
  \int\, d \Omega \,Y^m_\ell(\theta,\phi) =  \sqrt{4\pi}\delta^{m
  0}\delta_{\ell 0}, 
\end{align}

\noindent where $d \Omega = \sin(\theta) d\theta d\phi$, and
integrate Eq.~\eqref{1.3a} to obtain 
\begin{align}
  \R(0) = \int \frac{d^3k}{(2\pi)^3} \R(k) \exp{(\textrm{i}{\bf k\cdot r})}
  \vert_{r = 0} = &
  \int \frac{k^2  dk}{(2\pi)^3} \sum_{\ell = 0}^\infty
  \sum_{m = 0}^{\ell}  \sum_{n = 1}^\infty
  \R^m_{\ell|n} \left(\sqrt{4\pi}\delta_{\ell\,0}\delta_{m\,0}\right)
  \psi_n(k) , \nonumber\\
  = & \sum_{n = 1}^\infty  a^0_{0|n} \,\int \frac{dk}{\sqrt{\pi}(2\pi)^2}
  \psi_{n}(k) k^2  = \vartheta_0. \label{1.5}
\end{align}

\noindent To evaluate the central second derivative, we
follow the same steps and obtain
\begin{align}
    \R''(0) = &\int \frac{d^3k}{(2\pi)^3} \R(k)
    (\textrm{i} k)^2 \exp{(\textrm{i}{\bf k\cdot r})}\vert_{r = 0}
    \nonumber\\
    = & - \sum_{n = 1}^\infty \left(a^0_{0|n} +
    \sqrt{\frac{4}{5}}a^0_{2|n}\right) \int \frac{dk}{\sqrt{\pi}(2\pi)^2}
    \psi_{n}(k) k^4= \vartheta_2.\label{1.6}
\end{align}

\noindent The intermediate steps of the derivation of Eq.~\eqref{1.6}
are presented in Appendix \ref{appA}. 

 To proceed with the computation of the probabilities
given in Eqs. \eqref{delta:integral} and \eqref{delta:integralddr}
 we must integrate over all configurations in Fourier space. With the
 aid of the expansion \eqref{1.3b} we can express the measure of such
 integral in terms of the expansion coefficients satisfying the reality
conditions \eqref{1.4} and \eqref{1.41}, i.e., for any $\Psi[\R]$,
functional of $\R(\bk)$, the following integral can be represented as
\begin{align}
     \int\,\Psi[\R]\,[\d \R] =
     \Bigg[ \prod_{\ell =0}^{\infty} \prod_{m=1}^{\ell}\prod_{n=1}^{\infty}&
       \mu \int_{-\infty}^{\infty}\,\Psi[\R]\,\d a^m_{\ell|n}
     \int_{-\infty}^\infty \,\Psi[\R]\, \d b^m_{\ell|n}
     \Bigg]\times\notag\\
     \Bigg[&\prod_{\substack{p=0 }}^\infty \prod_{q = 1}^\infty
       \tilde{\mu} \int_{-\infty}^\infty\,\Psi[\R]\, \d a^0_{2p|q}
     \int_{-\infty}^\infty \,\Psi[\R]\,  \d b^0_{2p+1|q} \Bigg],\label{2.1}
\end{align}

\noindent where the constants $\mu$ and $\tilde{\mu}$ are weight
factors. In our calculation of probabilities, such factors are
absorbed by the final normalisation of the joint probability.

As mentioned before, we restrict ourselves to the Gaussian PDF. In
terms of the spherical harmonic coefficients (see Appendix
\ref{appB}), this means that 
\begin{align}
    \mathbb{P}[\R] = \exp\left( - \frac{1}{2 \pi^2(2\pi)^3}
    \sum_{\ell = 0}^\infty \sum_{m = 0}^{\ell}  \sum_{n = 1}^\infty
    |a^m_{\ell|n}|^2 + |b^m_{\ell|n}|^2  -
    \frac{1}{4 \pi^2(2\pi)^3}
    \sum_{p = 0}^\infty \sum_{q = 1}^\infty
    |a^0_{2p|q}|^2 + |b^0_{2p + 1|q}|^2 \right). \label{2.2}
\end{align}

\noindent In order to obtain the probabilities of the mentioned
parameters from Eqs. \eqref{delta:integral} and
\eqref{delta:integralddr}, we use the representation of the Dirac
$\delta$-function 
\begin{align}
  \delta(x) = \int^{\infty}_{-\infty} \,dz \,\exp[\textrm{i} z\,x].
\end{align}

\noindent  This allows us to write,
for example, the $\delta$-function in Eq.~\eqref{delta:integral} in
terms of the spherical harmonic coefficients as
\begin{align}
  \delta\left(\, \R(0) - \vartheta_0\, \right) =  \int \,dz
    \exp\left[\textrm{i} z\left( \sum_{n =1}^{\infty}a_{0|n}^{0}
    \int_0^\Lambda \, dk \psi_n\, k^2 
    - \frac{(2\pi)^3}{\sqrt{4\pi}} 
    \vartheta_0\right) \right]. \label{2.3}
\end{align}

\noindent In the same way, the representation of $\delta\left[\R''(0) -
\vartheta_2\right]$ can be written in terms of harmonic coefficients
with the aid of Eq.~\eqref{1.6}.

We now have all the elements needed to derive the probability of
the parameters $\R(0)$ and $\R''(0)$. Substituting expressions \eqref{2.2} and
\eqref{2.3} in Eq.~\eqref{delta:integral}, we perform 
the functional integral with the aid of the decomposition
\eqref{2.1}. In this process we discard all the Gaussian integrals
because they contribute to the probability only with a multiplicative
constant which will be included in the final normalisation. On the
other hand, the Dirac
$\delta$-function contributes with exponential factors of $a_{0|n}^0$
to the integrals. The integrals of these parameters are
computed by completing squares of the exponential arguments, so the 
integrals of such coefficients include a set of shifted Gaussian
functions (see Appendix~\ref{appB} for the details of this
procedure). The integral \eqref{delta:integralddr} can be performed
following the same steps and using the corresponding expressions
\eqref{1.6},\eqref{2.1} and \eqref{2.2}. The final probability density
for the pair of parameters $\R(0)$ and $\R''(0)$, is the product of
the integrals \eqref{1.5} and \eqref{1.6}, i.e.
\begin{align}
    \mathbb{P}\left(\,\R(0) = \vartheta_0,\, \R''(0) =
    \vartheta_2  \,\right) =  A
    \,\exp\left(-\frac{\vartheta_0^2}{2\Sigma_{(2)}^2}
    -\frac{5\,\vartheta_2^2}{2\Sigma_{(4)}^2} \right), \label{2.4}  
\end{align}

\noindent where $\Sigma_{(2)}$ and $\Sigma_{(4)}$ are the dispersion
of the amplitude and the second derivative respectively, and $A$
is a normalisation factor obtained from the condition that the
integral of the joint PDF over all possible values of the two
independent parameters equals unity.  The final normalised joint
probability density is 
\begin{align}
    \mathbb{P}(\vartheta_0,\vartheta_2) = \frac{4\sqrt{12}}{2\pi}
    \Sigma_{(2)}^{-1}\Sigma_{(4)}^{-1}
    \exp\left(-\frac{\vartheta_0^2}{2\Sigma_{(2)}^2}
    -\frac{5\,\vartheta_2^2}{2\Sigma_{(4)}^2} \right). \label{2.5}
\end{align}

\noindent According to the Press-Schechter formalism of structure formation
\cite{press74}, the PDF is integrated over all
perturbations which collapse to form  the astrophysical objects under
consideration. In this way we calculate the mass fraction of the
universe in the form of such objects. To apply this formalism and calculate the probability of PBH formation and integrate the
PDF \eqref{2.5}, we require the range of values  $\R(0)$ and $\R''(0)$
which correspond to PBH formation. In the next section we will obtain
this range with the help of the results of numerical computations
presented in \cite{musco07}. 

\section{The link between perturbation parameters and
  the curvature profiles used in numerical calculations}\label{metrics}

\subsection{Initial conditions}

As demonstrated by the first numerical simulations of PBH formation
\cite{nadezhin78}, whether or not an initial configuration with
given curvature profile leads to PBH formation, predominantly
depends on the following two factors:

1) The ratio of the size of the initial configuration $r_0$ to the size
   of the closed universe $r_{\mathsf{k}} = a(t)\int_0^1\, dr /
   \sqrt{1 - r^2}$ (evaluated at the initial time), which is a measure
   of the strength of gravitational field  within the
   configuration. 

2) The smoothness of the transition from the region of high curvature
   to the spatially flat FRW universe, which is 
   characterised by the width of the transition region at the edge of
   the initial configuration and it is inversely proportional to the
   pressure gradients there. Strong pressure
   gradients  inhibit PBH formation.

The numerical computations presented in \cite{musco07} (hereafter
 PM) give the time evolution of the configurations with initial
 curvature profiles  accounting for the 
above-mentioned factors and collapsing in a radiation-dominated
 universe. In that paper the initial conditions are obtained 
with the help of the quasi-homogeneous asymptotic solution valid in the
limit $t \to 0$. This solution to the Einstein equations was first
introduced by Lifshitz and Khalatnikov \cite{lifshitz63} (see also
\cite{zeldovich83,landau94}). Following   \cite{nadezhin78},  PM used this
asymptotic solution to set self-consistent initial
conditions for curvature
 inhomogeneities, the initial curvature
 inhomogeneity being described  by the spherically symmetric curvature
 profile $\K(\hr)$. This sets the 
initial conditions for the process of black hole
formation. Asymptotically, the metric can be presented in terms of
$\K(\hr)$ as 
\begin{align}
  ds^2 = s^2(\eta)\left\{- d\eta^2 + \frac{1}{1 -  \K(\hr)\hr^2}d \hr^2
    + \hr^2 \,\left[ d\theta^2 + \sin^2\theta d \phi^2 \right]\right\},  \label{4.2}
\end{align}

\noindent where $s(\eta)$ is the scale factor, $\eta$
is the conformal time and we
write $\hr$ for the radial coordinate to distinguish it from the
coordinate of the metric \eqref{0.1}. An advantage of working with
this metric is that it contains the curvature profile $\K(\hr) $
explicitly. We choose a set of coordinates with the origin at the 
centre of spherical symmetry and fix $\K(0) = 1$. The condition that
$\K(\hr)$ is a local inhomogeneity requires that $\K(\hr)
=0$ for  radii $\hr$ larger than the scale $\hr_0$, where
the metric matches the homogeneous FRW background.

In PM the profiles $\K(\hr)$ are presented in two forms,
one of which is characterised by two independent parameters $\alpha$ and
$\Delta$ as
\begin{align}
  \K(\hr)= \left[1 + \alpha \frac{\hr^2}{2 \Delta^2}\right] \exp\left(
  -\frac{\hr^2}{2 \Delta^2}\right). \label{K:profile}
\end{align}

\begin{figure}[htb]
 \centering
 \includegraphics[width = 0.45\textwidth]{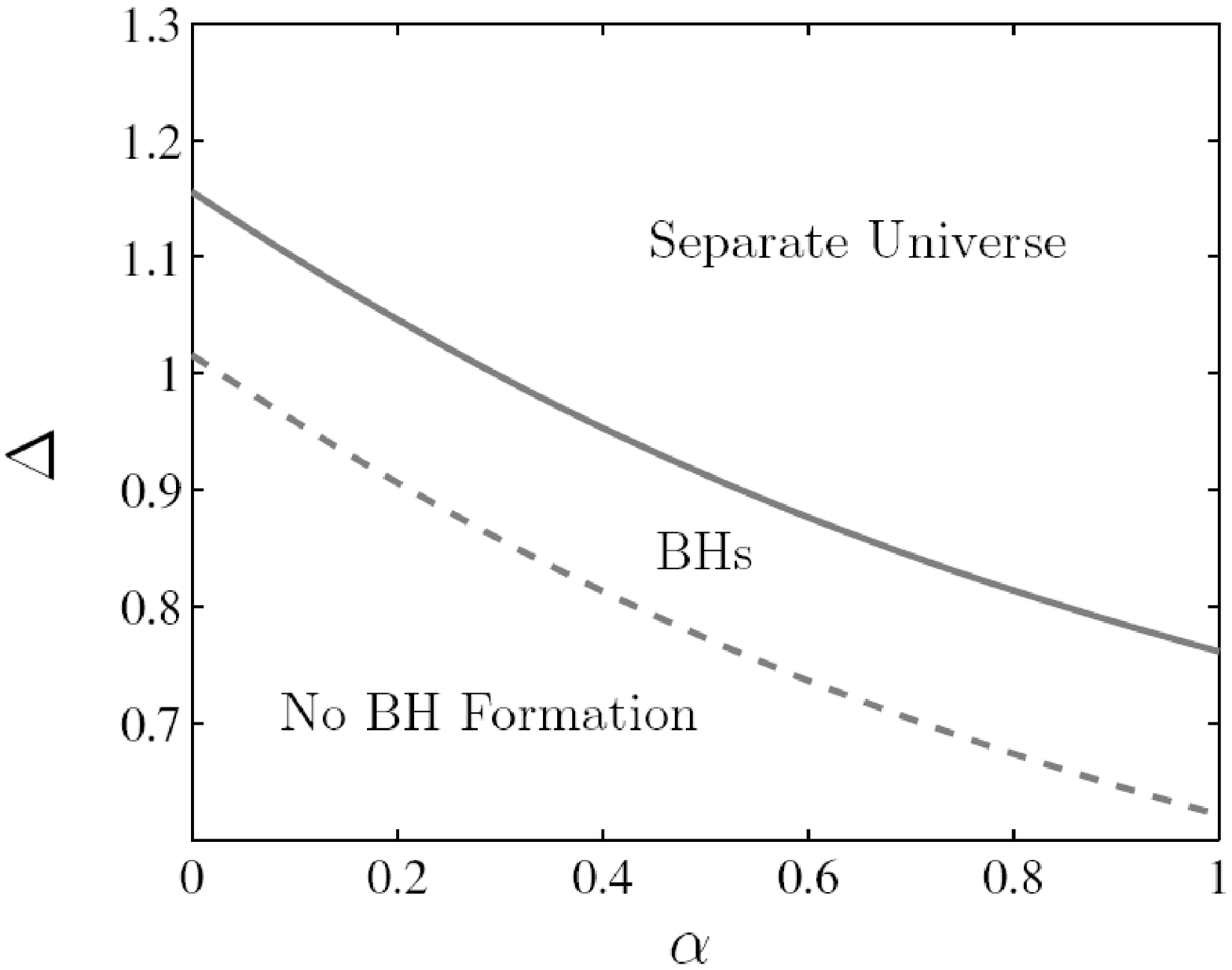}
 \hfill
 \includegraphics[width = 0.45\textwidth]{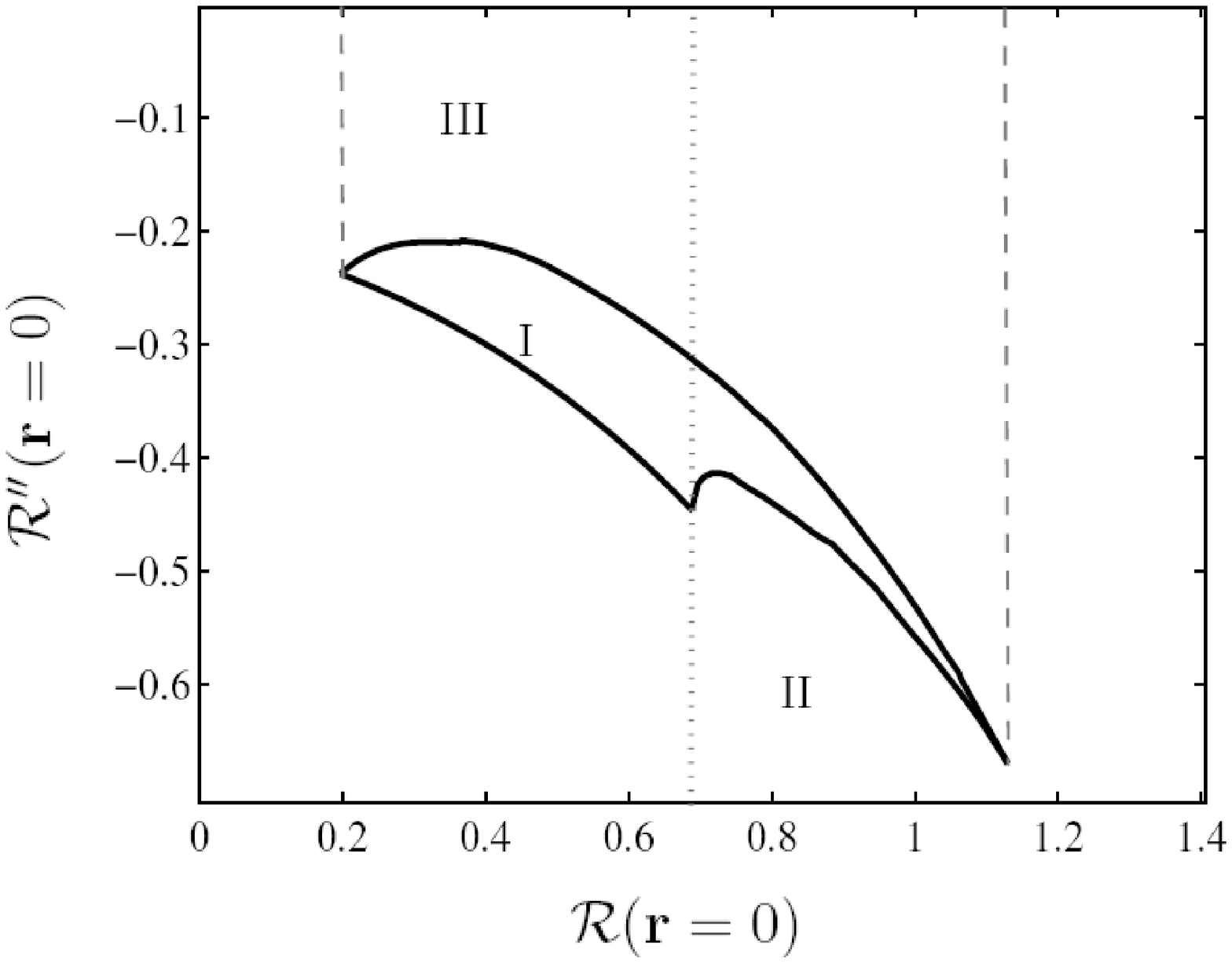}
 \caption{a) The left plot shows the parameter values for initial
 configurations which collapse to form black holes. $\Delta$
 characterises the width of the Gaussian curvature profile, while
 $\alpha$ characterises the deviations from a Gaussian profile, as can
 be seen in Eq. \eqref{K:profile}. b) In the
 $\left[\R(0),\R''(0)\right]$ plane three regions of integration are
 considered to compute the probability of PBH formation. Region I is
 the region enclosed by the solid curves and corresponds to the region
 noted by BH in Fig.~\ref{fig1}a. Region II is the region to the right
 of the grey dotted line representing the surface of integration
 considered in previous studies where only the amplitude is taken into
 account \cite{green04}. Region III is the region above the solid line
 and between the dashed lines. The physical characteristics of
 profiles with values in this region are described in section
 \ref{3c}.} \label{fig1}  
\end{figure}

\noindent The results of the numerical simulations in PM indicate that PBHs are
formed in the region of the parameter space $\left[\alpha,\Delta
  \right]$ shown in Fig.~\ref{fig1}a.

\subsection{Physical criteria for the identification of parameters }

We proceed by finding the correspondence between the two sets of
parameters, $[\R(0),\R''(0)]$ and $[\alpha,\Delta]$, both of which
describe the initial curvature profiles. Assuming that the
size of the configuration, $r_0$, is much larger than the Hubble
horizon $r_{H} = H^{-1}$, where $H$ is the 
Hubble parameter, we can use the gradient expansion of the
functions in metrics \eqref{0.1} and \eqref{4.2}. In this case, the
time derivative of any function $f(t,r)$ is of order $f / t \sim
Hf$ and significantly exceeds the spatial gradient which is of
order $f / r_0$. Hence the small parameter in the gradient expansion
is
\begin{align}
  \epsilon  \,\equiv \,\frac{r_H}{r_0}  = \frac{k}{a H}, \label{epsilon:def}
\end{align}

\noindent where $k$ is the wave-number corresponding to the scale of
the configuration. Taking into account that $\epsilon \to 0$ when $t
\to 0$, one sees that the gradient expansion is very
similar to the quasi-homogeneous solution \cite{lifshitz63}.

For the metric \eqref{0.1}, using the coordinate freedom to set $N^i =
0$ and ignoring any tensor contributions, i.e., taking $\tilde\gamma_{ij} =
\delta_{ij}$, the expansion of the Einstein equation
$G^{~0}_{0}~=~8\pi G T^{~0}_0$ to order $\epsilon^2$ can be written
as\footnotemark\footnotetext{For the complete second order expansion
  of the metric quantities, see for example,
  \cite{lyth05,langlois05}.}, 
\begin{align}
  \frac{1}{2}\left(\frac{6\dot{a}^2}{a^2}  + ^{(3)}\Rsp -
  \frac{4\dot{a}^2}{a^2} (N - 1) \right) + \mathcal{O}(\epsilon^{4})  =
  8\pi G\, (\rho_0 + \delta \rho) + \mathcal{O}(\epsilon^{4}),
  \label{einstein:00}
\end{align}

\noindent where $^{(3)}\Rsp$ is the spatial curvature, or the Ricci
scalar for the spatial metric $g_{ij}$. To order zero in $\epsilon$, we have
\begin{align}
   \frac{3 \dot{a}^2}{a^2} = 8\pi G\,\rho_0, \label{einstein:hom}
\end{align}

\noindent which corresponds to the homogeneous part of
\eqref{einstein:00}. As shown in \cite{shibata95,shibata99,tanaka06},
the time slicing can be set to a uniform expansion gauge in which
\begin{align}
  N - 1 =  - \frac{3 \Gamma - 2}{\Gamma} \delta + 
  \mathcal{O}(\epsilon^4),\label{N:delta}
\end{align}

\noindent where $\Gamma - 1$ is the sound-speed squared. Using
\eqref{einstein:00},\eqref{einstein:hom} and \eqref{N:delta}, we find
the equivalence between the spatial curvature and the matter overdensity
\begin{align}
  ^{(3)}\Rsp = \frac{8 \pi G}{3}\delta \rho  \,\left(\frac{4 +
  3\Gamma}{3\Gamma}\right).  \label{einstein:inhom} 
\end{align}

\noindent In consequence, the gradients establish a correspondence
with the pressure gradients
\begin{align}
  \nabla {^{(3)}\Rsp} = \frac{8 \pi G}{3} \,\frac{4 +
  3\Gamma}{3\Gamma} \nabla\left(\delta \rho\right)  = \frac{8 \pi G}{3} 
  \,\left(\frac{4 + 3\Gamma}{3\Gamma\left(\Gamma - 1\right)}\right) \nabla p,
  \label{einstein:grad} 
\end{align}

\noindent where $\nabla = (g_{rr})^{-1/2} \d / \d r$. The last equation
shows that the gradient of the spatial curvature is
directly related to the pressure gradient. Hence, subject to these two
physical conditions at the edge of the configuration, we
relate the profiles $\R(r)$ and $\K(\hr)$ by equating the spatial
curvature and its gradient for metrics \eqref{0.1} and
\eqref{4.2}. That is,  
\begin{align}
   ^{(3)}{\Rsp} =  - \left[2 \R''(r) + \left(\R'(r)\right)^2
  \right] \exp(-2 \R(r)) = 3\K(\hr) + \hr \K'(\hr), \label{3curv:eq}
\end{align}

\noindent and
\begin{align}
  \frac{1}{\sqrt{\mathrm{g}_{rr}}}\frac{\d }{\d r} &\left( ^{(3)}\Rsp
  \right) = \notag\\
   - \left[ \R'\R'' + \R''' \right]& \exp(-3 \R(r)) =
   \left[\frac{1 - \K \hr^2}{\hr^2}\right]^{1/2}
   \left(2 \hr \K'(\hr) + \frac{1}{2}\hr^2 \K''(\hr)\right). \label{3curv:der}
\end{align}

\noindent By definition of the edge of curvature configuration, the three
curvature must vanish at this point, so Eq.~\eqref{3curv:eq} implies
\begin{align}
   2 \R''(r_0) + \left(\R'(r_0)\right)^2 = 0, \label{3curv:r-0}
\end{align}
and
\begin{align}
   3\K(\hr_0) + \hr_0 \K'(\hr_0) = 0. \label{3curv:hr-0}
\end{align}

\noindent As a consequence of this, the gradient relation
\eqref{3curv:der} can be written as
\begin{align}
    \left[\R'(r_0)^3 - 2 \R'''(r_0) \right]& \exp(-3 \R(r_0)) =
   \left[\frac{1 - \K \hr_0^2}{\hr_0^2}\right]^{1/2}
   [-12 \K(\hr_0) + \hr_0^2 \K''(\hr_0)]. \label{3curv:der-0}
\end{align}

\noindent This establishes a relation between $\R(r)$ and $\K(\hr)$
at the edge points $r_0$ and $\hr_0$. The configuration $\K(\hr)$ is
parameterised by $[\alpha, \Delta]$, as shown in
Eq.~\eqref{K:profile}. As follows from condition \eqref{3curv:hr-0}
(see also PM), the radius $r_0$ can be written in terms of those
parameters as 
\begin{align}
  \hr_0^2 =\left( \frac{5 \alpha  -  2 + \sqrt{(5\alpha - 2)^2 -
      24\alpha}}{2\alpha} \right)\Delta^2. \label{rhat:0}
\end{align}

\noindent  Then we use two more equations obtained from the conformal
transformation of coordinates at zero order in
$\epsilon$:
\begin{align}
  a^2(\tau) \e{2\R(r)}\,dr^2 = s^2(\eta)\frac{d\hr^2}{{1 -
      \K(\hr)\hr^2}}
  \label{radial:elem}
\end{align}
\noindent and
\begin{align}
    a^2(\tau)\, \e{2\R(r)}\, r^2\, d\Omega^2 = s^2(\eta)\, \hr^2\, d\Omega^2.
\end{align}

\noindent Because the homogeneous Einstein equations are identical
in both metrics, the scale factors $a(\tau)$ and $s(\eta)$ can be
identified, $a(\tau) \equiv s(\eta)$. Thus we find a relation between
the radial coordinates,
\begin{align}
 \e{\R(r)}\, r =  \hr \label{r:hr}
\end{align}

\noindent and an integral relation between the configurations
\begin{align}
   \int_0^r \e{\R(x)}\,dx =  \int_0^{\hr} \frac{dx}{\sqrt{1 -
   \K(x)x^2}}.
   \label{radial:integral}
\end{align}

\noindent One can verify that Eqs.
\eqref{3curv:eq}, \eqref{3curv:der} and \eqref{radial:integral} are
not independent. For example, Eq.~\eqref{3curv:der} follows from
\eqref{3curv:eq} and \eqref{radial:integral}. 

In the previous section we have developed a method to account
for the probability of any set of parameters describing the curvature
profile. For simplicity we have chosen the pair $[\R(0),\R''(0)]$. We
now illustrate how to relate $[\R(0),\R''(0)]$ and $[\alpha, \Delta]$
by considering the parabolic profiles
\begin{align}
  \R(r) =\R(0) + \frac{1}{2}\R''(0)\,r^2. \label{param:expansion}
\end{align}

\noindent This parametrisation meets the minimal requirement of
covering the $[\alpha, \Delta]$  parameter space  in Fig.~\ref{fig1}a.
\begin{figure}
  \begin{center}
    \includegraphics[totalheight=0.30\textheight]{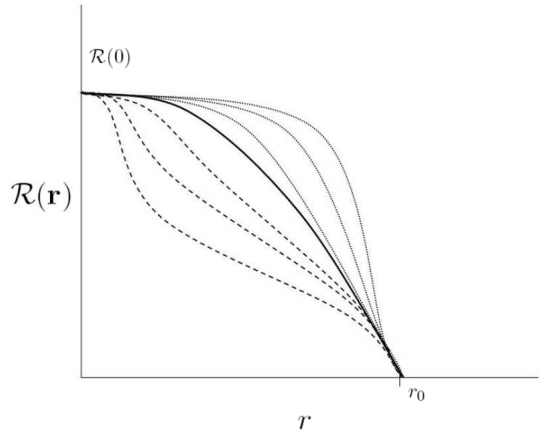}
    \caption{The curvature profile for three different families of
      configurations with common central amplitude $\R(0)=1$. The
    configurations shown by the dashed lines have value of $\R''(0) $
    larger in absolute magnitude than the parabolic one shown in
    black. The configurations shown by the dotted lines have a value
    of $\R''(0)$ smaller than the parabolic one. All profiles satisfy
    conditions \eqref{3curv:r-0}  and 
    \eqref{3curv:der-0}.}\label{fig11} 
  \end{center}
\end{figure}

\noindent Eqs. \eqref{3curv:r-0}, \eqref{radial:integral} and \eqref{r:hr}
are now reduced to the following system of algebraic equations:
\begin{align}
  &r_0^2 = - \frac{2}{\R''(0)}, \label{x:0}\\
  \R(0) = 2 \log&\left(\frac{2}{\textrm{erf}(1)}\left[\pi \exp(1) \hr_0
    \right]^{-1/2} \int_{0}^{\hr_{0}} \frac{dx}{\left(1  -
    \K(x)x^2\right)^{1/2}}\right), \label{R:centre}\\
  \R''&(0) = - 2 \frac{\exp(2 \R(0) - 2)}{\hr_0^2},\label{ddR:centre}
\end{align}
where $\hr_0$ is given in terms of $[\alpha, \Delta]$ by
Eq.~\eqref{rhat:0}. 

\begin{figure}[hbt]
  \begin{center}
    \includegraphics[width = 0.45\textwidth]{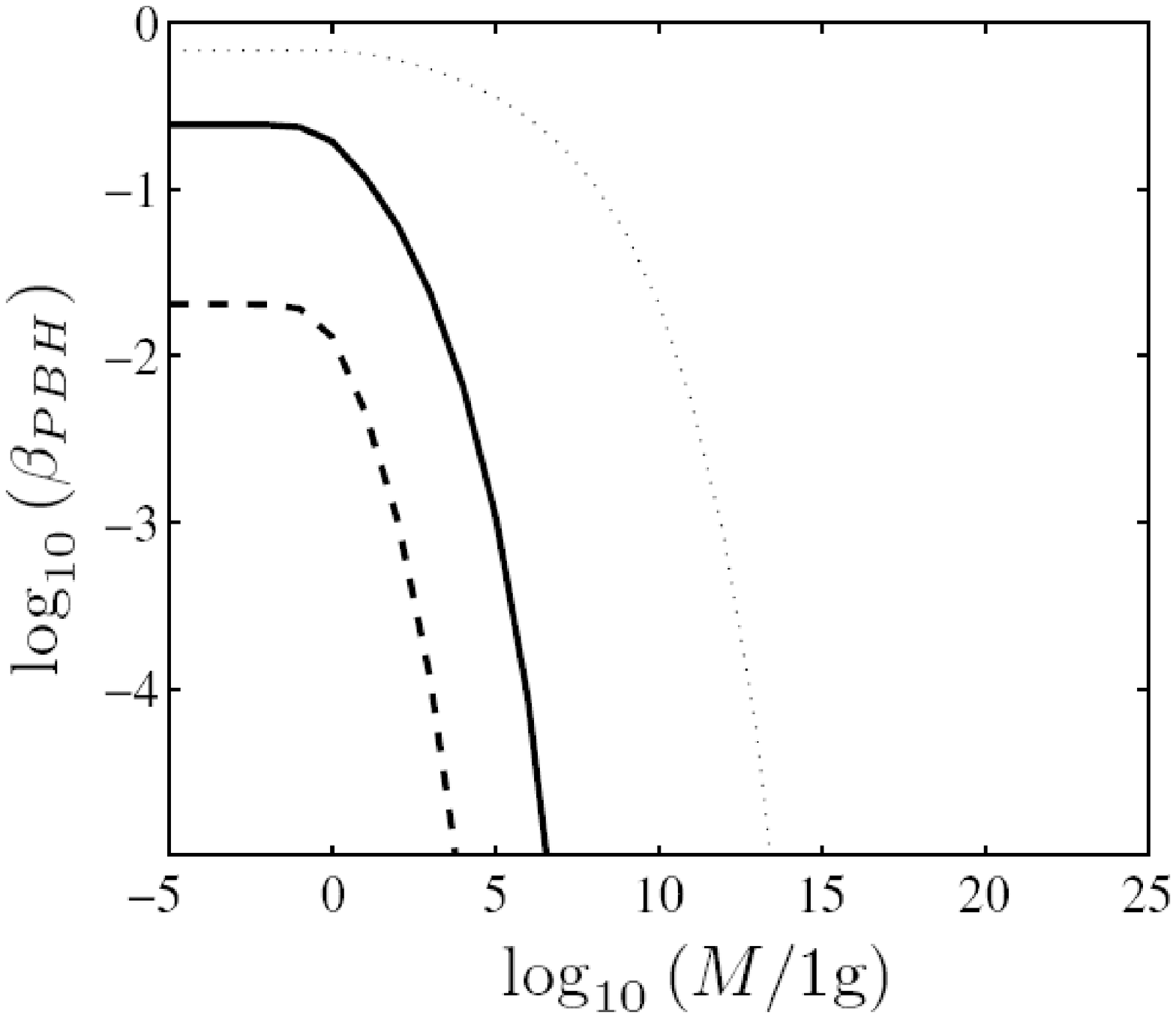}
    \hfill
    \includegraphics[width = 0.45\textwidth]{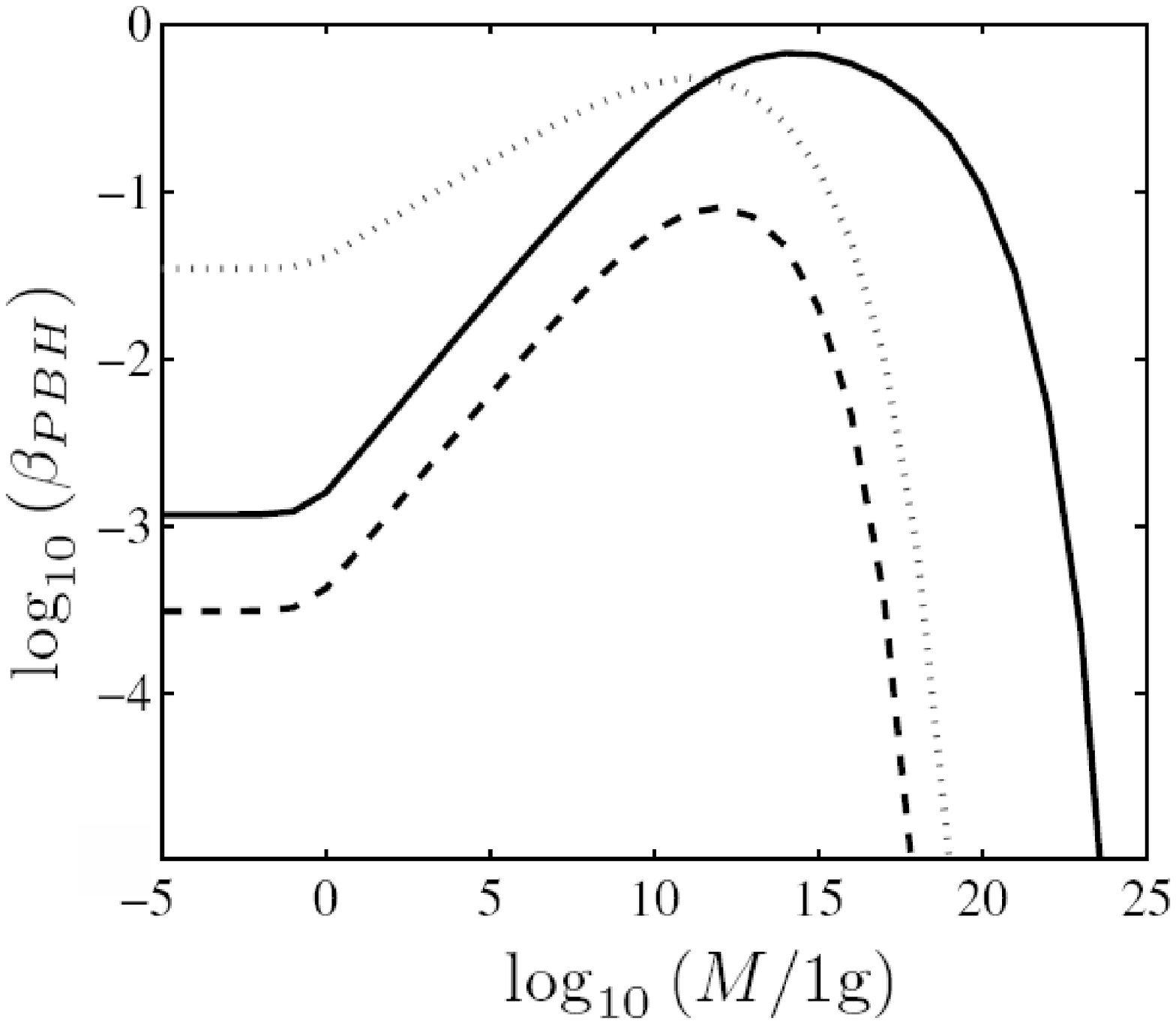}
    \caption{The logarithmic probability of PBHs $\beta$
    calculated using Eq. (42) with a power spectrum with two tilt
    values ($n_s = 1.32$ for the left plot, $n_s = 1.47$ for the right
    plot). The lines show the integration for the three different
    regions sketched in Fig.~1. The integral over the region I
    ($\beta_{\rm I}(M)$) corresponds to the dashed lines, and the
    integration over the region II ($\beta_{\rm II} (M)$) to the solid
    lines. The probability integrated over the region III ($\beta_{\rm
    III}(M)$) is represented by the dotted lines in both figures.}  
    \label{fig2} 
  \end{center}
\end{figure}

\subsection{Parameter values leading to PBH formation}
\label{3c}

As follows from the numerical computations \cite{musco07}
which used the parametrisation \eqref{K:profile}, PBHs are formed in
the [$\alpha$, $\Delta$] region shown in
Fig.~\ref{fig1}a. Equations \eqref{R:centre} and
\eqref{ddR:centre} map this region to the Region I in the space of
parameters [$\R(0)$, $\R''(0)$] shown in Fig.~\ref{fig1}b. The
Jacobian of the transformation corresponding to this mapping is
non-vanishing, which guarantees a one-to-one correspondence of the
region 'BHs' plotted in Fig.~\ref{fig1}a with Region I in
Fig.~\ref{fig1}b. Each point here corresponds to a parabolic profile
which leads to the formation of PBH. 

For each one of these parabolic profiles, there is a family of
non-parabolic profiles with the same central amplitude $\R(0)$, the
same configuration size $r_0$, and the same behaviour near
the edge, as shown in Fig.~\ref{fig11}. In that
figure, the profiles lying below the parabola correspond to larger
absolute magnitudes of $\R''(0)$ and do not form PBHs because they
have lower average  gravitational field strength and
higher average pressure gradient. The non-parabolic
profiles which lie above the parabolic one (with smaller absolute
magnitude $\R''(0)$) should also collapse to form PBHs because they
correspond to higher average gravitational field strength and
lower pressure gradient. 

In the  parameter-space  $[\R(0),\, \R''(0)]$, this last set of
profiles  corresponds to Region III in Fig.~\ref{fig1}b. This region
will be included in the calculation of the probability of PBH
formation in the next section.

 \section{Two parametric Probability of PBH formation}
    \label{prob:comparison}

To calculate the probability of PBH
formation, which is equivalent to the mass fraction of the universe
going to PBHs of given mass, it is customary to use the standard
Press-Schechter formalism \cite{press74}. This has been widely used in
previous calculations of the one parametric probability of PBH
formation \cite{carr75, lidsey94,liddle98, carr05,siri07,zaballa06}.
When the probability depends on a single amplitude
parameter, this method  reduces to  the integration of the
corresponding PDF over the relevant perturbation amplitudes. The 
final integral is equivalent to the mass fraction of PBHs of
mass $M \sim (\Gamma - 1)^{3 / 2} M_{H} \approx (\Gamma -1)^{3 / 2} k_M /
(2\pi)$ \cite{carr75}, with the soundspeed $\sqrt{\Gamma - 1}$
measured at the time formation\footnotemark\footnotetext{Throughout
  this paper we consider configurations that collapse in a uniform
  radiation dominated background. Thus we use the value $\Gamma  = 4/3$}. 
Here we extend the standard Press-Schechter formalism to derive a two
parametric probability, introducing  the second derivative at the
centre of the configuration as an additional parameter. When
the  $\left[\R'(0),\,\R''(0)\right]$ region is a square
$[\R_1<\R(0)<\R_2$, $\R''_1<\R''(0)<\R''_2]$, the integrated two parametric
probability for objects of mass $M$ is 
\begin{align}
  \beta_{PBH}(M) = \,\int_{\R_{\rm 1}}^{\R_{\rm 2}} 
  \,d\vartheta_0 \int_{\R''_{\rm 1}}^{\R''_{\rm 2}}  \, d \vartheta_2 \,
  \mathbb{P}(\vartheta_0,\vartheta_2) = \qquad \qquad\qquad \qquad&\notag \\
  \frac{1}{2}\left[{\rm erf}\left(\frac{\R_{\rm
	2}}{\sqrt{2}\Sigma_{(2)}(M)}\right)-  {\rm
      erf}\left(\frac{\R_{\rm
	1}}{\sqrt{2}\Sigma_{(2)}(M)}\right)\right]& \times
  \left[{\rm erf}\left(\frac{\R''_{\rm 2}}{\sqrt{2}\Sigma_{(4)}(M)}\right)
    - {\rm erf}\left(\frac{\R''_{\rm 1}}{\sqrt{2}\Sigma_{(4)}(M)}\right)
    \right].
  \label{proba:integral}
\end{align}

\noindent We use this result to integrate numerically over a mesh of
small squares covering each one of the regions of the plane
$\left[\R(0),\,\R''(0)\right]$ shown in Fig.~\ref{fig1}b. We call the 
integral over region I $\beta_{\rm I}(M)$, and correspondingly the
integrals over regions II and III are called  $\beta_{\rm II}(M)$ and
$\beta_{\rm III}(M)$. The mass dependence of these betas for two
different power-law spectra  $\ps_{\R}(k) \propto k^{n-1}$, are
shown in Fig.~\ref{fig2}. As dictated by the Press-Shechter
  formulation, such integration corresponds to the fraction of mass
  density in the universe  that has collapsed into objects with mass
  $M$. We remind the reader that our choice of $n_s$ and the
mean amplitude is for pure illustration purposes. With the values
used in this paper, copious amounts of black holes are produced. A red
spectral index and power spectrum inferred from the CMB data
corresponds to a low number of PBHs. However, to assume that the same
values of the power spectrum and spectral index are valid on scales
relevant to PBH formation (30 decades of mass below the mass scales
correspondent to CMB observations) is a very strong extrapolation. At
the present time we cannot exclude that the values of $n_s$ and the
power spectrum are different than those given by the CMB.  To explore
the structure formation models that match CMB observation values and
also produce considerable number of PBHs is a great task beyond the
scope of our paper. This important issue is currently under
investigation \cite{kohri07,bugaev08}.

\begin{figure}[htb]
  \begin{center}
    \includegraphics[width = 0.55\textwidth]{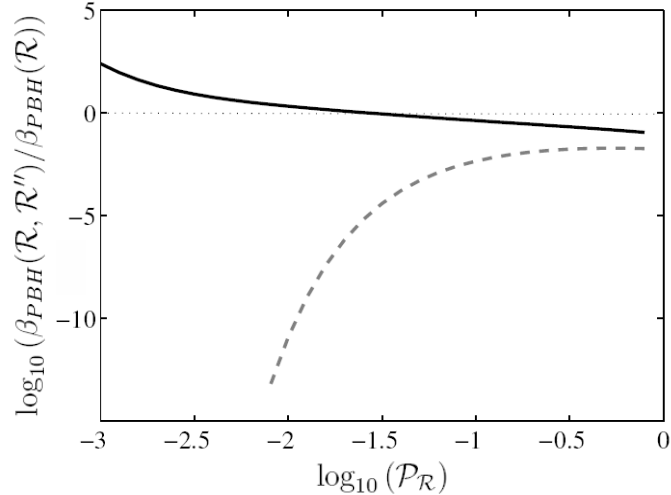}
    \caption{The horizontal axis of the figure is the
	amplitude of the power spectrum at scales relevant to PBH
	formation. The grey dashed line shows the ratio
	$\beta_{\rm I}/\beta_{\rm II}$ where $\beta_{\rm I}$ is the probability
	density integrated over region I in the
	$[\R(0),\R''(0)]$-parameter space of Fig.~1b, and $\beta_{\rm II}$
	is the corresponding probability integrated over region II of
	the same figure. The black line is the ratio $\beta_{\rm III}$
	over $\beta_{\rm II}$. The dotted line shows the reference case of
	the single-parameter probability.}
      \label{fig3}
  \end{center}
\end{figure}

We contrast the case of parabolic profiles described by
Eq.~\eqref{param:expansion} with the non-parabolic set presented in
Fig.~\ref{fig11} by plotting the ratios of probabilities $\beta_{\rm
  I}/ \beta_{\rm II}$ and $\beta_{\rm III} / \beta_{\rm I}$ for
different values of $\ps_{\R}$.  This is presented in
Fig.~\ref{fig3}. This figure shows that the probability of PBH
formation can be larger than the previous one-parameter probability
computed from the integration of Region II as done in previous studies
\cite{green04}. This important result requires confirmation from more
detailed numerical simulations of PBH formation in this region of
parameter-space. The uncertainty is explained by the fact that the two
parametric calculation of the probability of PBH formation is still
incomplete. This should be complemented in the future by the
introduction of all relevant higher derivative parameters and the
higher-order correlations in the PDF.

\section{Discussion}\label{conclusion}

We have developed a method for
calculating the two-parametric probability of PBH formation, taking
into account  the radial profiles of non-linear curvature
cosmological inhomogeneities. This is the fist step towards
calculating the $N$-parametric probability, which takes into account
the radial profiles more precisely than studies using the amplitude as
the only relevant parameter.  Using the results of sophisticated
numerical computations, we obtain the values of $\R''(0)$ that are
relevant for PBH formation. Subsequently we have incorporated these
values to the total probability of PBH formation. Finally, we have
provided an example of the consequences of this probability for the
statistics of PBHs.

The results obtained show that, if we restrict ourselves to the PBH
formation calculated for parabolic profiles only (as described in
Section \ref{metrics}), then the total probability of PBH formation is
orders of magnitude below  previous estimations. On the other hand, 
with the aid of heuristic arguments we show that a much larger region
of parameter-space $[\R(0),\R''(0)]$ representing non-parabolic
profiles should also be considered in the estimation of the
probability of PBH formation (see Fig.~\ref{fig11}). In this case,
the total probability of PBH formation is higher than the
single-parameter estimate of previous works. In this
case, we have an opportunity to impose new bounds on the power
spectrum on the scales relevant for PBH formation. Analysing the
uncertainty of our results, mostly due to the heuristic nature
  of the present study, we have demonstrated how much we still 
have to understand about the formation and statistics of PBHs. The
physical arguments supporting our results should be made rigorous by
direct verification with numerical hydrodynamical simulations of PBH
formation. This in turn would provide  valuable support for
the initial motivation of this work.  

The main conclusion of this paper is that the amplitude of
initial inhomogeneities is not the only parameter which determines the
probability of PBH formation. The ultimate solution of the problem
requires a greater set of parameters and a larger range of their
values to determine all high curvature configurations that form
PBHs, which is a huge task for future research. In the meantime,
we have a method to operate with the statistics of all these
parameters.

  \section{Acknowledgements}

JCH gratefully acknowledges financial support from the Mexican
Council for Science and Technology (CONACYT) (Studentship
No.~179026). We would like to thank Prof~Bernard~Carr, Dr~Karim~Malik,
Dr~David~Seery and Dr~Ilia~Musco for useful comments and discussions.

\appendix


\section{Harmonic decomposition of $\R$ and Fourier representation of
  $\R(0)$ and $\R''(0)$}\label{appA}

The Fourier expansion of the smoothed curvature perturbation profile
$\R$ in terms of spherical harmonic functions is,
\begin{align}
  \R(\bk,t) = &\sum_{\ell = 0}^\infty\,
    \sum_{m = - \ell}^{\ell}\,  \sum_{n = 1}^\infty\,
    \R^m_{\ell|n}(t) Y_{\ell m}(\thetak,\phik)
    \psi_n(k). \label{fourier:expansion}
\end{align}

\noindent The radial functions of the harmonic decomposition can be
defined by
\begin{equation}
    \label{harmonic:psi}
    \psi_n(k) = \frac{\sqrt{2}}{J_{\nu+1}(\alpha_\nu^n)} \frac{\ps(k)
    \window(k;k_\mathrm{M})}
    {\Lambda k^2} J_\nu\left( \alpha_\nu^n \frac{k}{\Lambda}
    \right),
\end{equation}

\noindent where $\window(k;k_\mathrm{M})$ is the window function with smoothing
scale $k_\mathrm{M}$ and $\alpha_\nu^n$ is the $n$-th root of the Bessel
function of order $\nu$, $J_{\nu}(k)$. Note that the functions of the
radial coordinate in the expansion are the Bessel functions up to a factor.
 The set of functions $\psi_n(k)$ is orthonormal under the product
\begin{equation}
  \label{harmonic:orthonormal}
  \int_0^\Lambda \d k \; \frac{k^5}{\ps(k) \window^2(k;k_\mathrm{M})}
  \psi_m(k) \psi_n(k) = \delta_{m\,n},
\end{equation}

\noindent and the completeness relation can be written with the aid of
the Kernel in the internal product: 
\begin{equation}
  \frac {k_0^5}{\ps(k_0) \window^2(k_\mathrm{M})}\sum_n \psi_n(k_0) \psi_n(k)
  = \diracd(k-k_0).\label{harmonic:complete}
\end{equation}

\noindent In Eq.~\eqref{harmonic:orthonormal} $\Lambda$ represents an
artificial compactification of the momentum space which is used only
to have an explicit definition of $\psi$ at hand. With the definitions
above, the expansion \eqref{fourier:expansion} describes the curvature
perturbation with power spectrum $\ps(k)$ smoothed over a scale
$k_\mathrm{M}$. 

Let us now construct a parameter to represent the second radial
derivative of the field $\R$ in Fourier space. The first radial
derivative of $\R(\br)$ is   
  \begin{align}
    \frac{\partial}{\partial r} \R(\br) = \int \,d^3 k \sum_{l,m,n}
    \R^{m}_{l,n} Y^{m}_{l}(\thetak,\phik)
    \psi_{n}(k)\e{\mathrm{i}(\bk\cdot\br)}\times
    \frac{\partial}{\partial r} (\mathrm{i}
    \bk\cdot\br).\label{R:der1} 
  \end{align}

\noindent If we write the spherical coordinates in a Cartesian basis, we have
\begin{align}
  \bk\cdot \br = |k||r|
  \left\{\sin(\theta_{r})\cos(\phi_{r})\sin(\thetak)\cos(\phik)  +
  \sin(\theta_{r})\sin(\phi_r)\sin(\thetak)\sin(\phik) +
  \cos(\theta_{r})\cos(\thetak) \right\}, \label{kx:product}
\end{align}

\noindent with $(\phi_{r},\theta_{r})$ the set of angles of
the vector $\br$ and $(\phik,\thetak)$ the corresponding pair for
$\bk$.

The scalar product \eqref{kx:product} can be simplified if we note
that the integral in Fourier space spans all possible directions of
$\bk$, so we can choose an arbitrary direction for $\br$. In
particular, we can fix the $\br$ so that
$\cos(\theta_{r}) = 1$. This allows us to write a simple expression for the
radial derivative,
\begin{align}
  \frac{\partial}{\partial r} (\bk\cdot\br) = |k| \cos(\thetak).
\end{align}

\noindent The first derivative of the profile $\R(\br)$ at the centre of the
configuration is zero by the symmetry of spherical
configurations. By construction of the spherically symmetric Fourier modes, this
condition is satisfied identically at $\br = 0$. The first non-vanishing
parameter that gives information about of the profile of perturbations is the
second derivative. Fixing the direction of the vector $\br$ in
the scalar product we have 
\begin{align}
  \frac{\partial^2}{\partial r^2} \R(\br) = \int \,d^3 k \sum_{l,m,n}
  \R^{m}_{l,n} Y^{m}_{l}(\thetak,\phik) \psi_{n}(k)\times k^2
  \cos^2(\thetak) \e{\mathrm{i}(\bk\cdot\br)}.\label{R:der2}
\end{align}

\noindent  With the standard definition of the spherical harmonics,
\begin{align}
Y^{m}_{l}(\theta,\phi) = \sqrt{\frac{2l +1}{4\pi}
  \frac{(l-m)!}{(l +  m)!}}P^m_l
(\cos(\theta))\e{\mathrm{i}\,m\phi},  \label{sh:def}
\end{align}
\noindent where the normalisation factor is used for orthonormality
purposes, one can write the factor $\cos^2(\thetak)$ as the sum of
two spherical harmonics,
\begin{align}
  \cos^2(\thetak) = \frac{1}{3} \sqrt{4\pi}\left( \sqrt{\frac{4}{5}}
  Y^0_2 + Y^0_0\right) = \frac{1}{3} \sqrt{4\pi}\left( \sqrt{\frac{4}{5}}
  \bar{Y}^0_2 + \bar{Y}^0_0\right),\label{decomp:cossq}
\end{align}
\noindent where the $\bar{Y}$ indicates the complex conjugate.

Using the normalisation rule for spherical harmonics,
\begin{align}
  \int\, d \Omega \,Y^m_l(\theta,\phi)\bar{Y}^n_k(\theta,\phi) =
  \delta_{m\,n}\delta_{l\,k},\label{harmonic:completness}
\end{align}

\noindent  we can integrate Eq.~\eqref{decomp:cossq} in the derivative
\eqref{R:der2} and arrive at the expression
\begin{align}
\R''(0) = \int\, dk k^4 \left(\frac{\sqrt{4\pi}}{3}\right) \left[
\sum_{n=1}^{\infty} \left(\sqrt{\frac{4}{5}}\R_{2,n}^0 +
\R_{0,n}^0\right) \psi_n \right]. \label{R:biprima}
\end{align}

\noindent This is the result used in Section \ref{prob:construction}.

To finish this appendix we show how the introduction of a parameter
off the centre, say $\R'(r_0)$, generates a large set of constraints
on the values of the coefficients $\R_{\ell|n}^0$. The integral in
Fourier space representing this derivative is
\begin{align}
  \R'(r)|_{r = r_0} = & \mathrm{i}\,\int \,\frac{d^3 k}{(2\pi)^3}\,
  |k|\, \R(\bk) \cos(\theta_k) \exp\left[\mathrm{i}\, k \,r
    \cos(\theta_k)\right]\notag\\ 
  = &  \mathrm{i}\, \int \, \frac{d^3 k}{(2\pi)^3}|k|\,\R(\bk)
  \,Y^0_1(\theta,\phi) \left[\sum_{s = 0}^{\infty}
    \frac{(\mathrm{i}\cos(\theta_k) r_0 k)^s}{s!}\right] 
  \label{rprime:r0} 
\end{align}

\noindent where we have expanded the exponential function  in Taylor
series. Each power of $\cos(\theta)$ can be expressed in terms of
spherical harmonic functions $Y_{\ell}^0$. This means that the last
integral consists of a series of terms of the form, 
\begin{align}
  \int\,d\Omega \,
	 Y^n_\ell(\theta,\phi)Y^0_1(\theta,\phi)Y^0_S(\theta,\phi).
\end{align}

\noindent Each of these integrals is a Clebsch-Gordan
coefficient. These steps are enough to show that, while a parameter
$\R'(r = r_0)$ might represent an improvement in the estimate of the
final probability of PBH formation, it takes a long calculation to
complete squares, add normalisation factors for each coefficient
$\R^0_{\ell|n}$, and arrive at a final expression like
eq. \eqref{2.5}. This goes beyond the goals of the present paper.
    

\section{The probability of $\R(0)$ and  $\R''(0)$} \label{appB}

At any time $t$, the probability distribution $\Prob_t[\Rsp]$, is
formally obtained through the inverse Fourier transform of
$Z_t[\eta]$, a generating functional which can be expanded in terms of
the $n$-point correlation functions \cite{sh06},   
\begin{equation}
  \label{genfunc:reconstruct}
  Z_t[\eta] = \exp \sum_{n=0}^\infty \frac{\imag^n}{n!}
  \int \cdots \int \d^3 y_1 \cdots \d^3 y_n \; \eta(\by_1)
  \cdots \eta(\by_n) \langle \R(t,\by_1) \cdots \R(t,\by_n)
  \rangle ,
\end{equation}

\noindent Hence, up to an overall normalisation,
\begin{equation}
  \label{genfunc:prob}
  \Prob_t[\R] \propto \int [\d \eta] \; \exp\left( - \imag \int \d^3 \mathrm{x} \;
  \R(\bx)\eta(\bx) \right) Z_t[\eta] .
\end{equation}

\noindent    If all correlation functions of three and more points are set to
    zero, then $\Prob[\R] \propto \G[\R]$. Assuming this for the
    generating functional \eqref{genfunc:reconstruct}, the expression
    to integrate in the Fourier space is, 
\begin{equation}
  \Prob_t[\eta;\R] =
  \exp\left(- \int \frac{\d^3 k_1 \, \d^3 k_2}{(2\pi)^6} \;
  \frac{\eta(\vect{k}_1) \eta(\vect{k}_2)}{2}
  \langle \R(t,\vect{k}_1) \R(t,\vect{k}_2) \rangle - \imag \int
  \frac{\d^3 k}{(2\pi)^3} \;
  \eta(\vect{k}) \R(\vect{k}) \right) .
\end{equation}

\noindent The functional integral of this expression gives the probability of
$\R$. To solve this integral we  complete the square of $\eta$ factors
and make the finite field redefinition
\begin{equation}
  \eta(\vect{k}) \mapsto \hat{\eta}(\vect{k}) = \eta(\vect{k}) +
   \imag(2\pi)^3 \frac{\Rsp(\vect{k})}{\langle \R(t,\vect{k})
   \R(t,-\vect{k}) 
  \rangle'} ,
\end{equation}
where the prime $'$ attached to $\langle \R(t,\vect{k}) \R(t,-\vect{k})
\rangle'$ indicates that the momentum-conservation $\delta$-function
is omitted. The measure $[\d\eta]$ is invariant under this
shift, giving $\int [\d \eta] = \int [\d \hat{\eta}]$,
whereas $\Prob_t[\eta;\R]$ can be split into an $\R$-dependent piece,
which we call $\G_t[\R]$, and a piece that depends only on
$\hat{\eta}$ but not $\R$,  
\begin{equation}
  \Prob_t[\eta;\R] \mapsto
  \G_t[\R] \exp \left( - \frac{1}{2} \int \frac{\d^3 k_1 \, \d^3 k_2}
  {(2\pi)^6} \; \hat{\eta}(\vect{k}_1) \hat{\eta}(\vect{k}_2)
  \langle \R(t,\vect{k}_1) \R(t,\vect{k}_2) \rangle \right) ,
\end{equation}
where $\G_t[\R]$ is a Gaussian in $\R$,
\begin{equation}
  \G_t[\R] = \exp \left( - \frac{1}{2} \int \d^3 k_1 \, \d^3 k_2 \;
  \langle \R(t,\vect{k}_1) \R(t,\vect{k}_2) \rangle
  \frac{\R(\vect{k}_1) \R(\vect{k}_2)}{\prod_i
  \langle \R(t,\vect{k}_i) \R(t,-\vect{k}_i) \rangle'} \right) .
\end{equation}

\noindent When we make the expansion of the fields $\R(\bk)$ in terms of the
spherical harmonics as in Eq.~\eqref{fourier:expansion} and using the
explicit expression for the two point correlation
Eq.~\eqref{intro:variance} we obtain, 
\begin{eqnarray}
  \nonumber
  \G[\R] = \exp \Bigg( - \frac{1}{2} \int \d\Omega \int k^2 \, \d k \;
  \frac{k^3}{(2\pi)^3 2\pi^2} \frac{1}{\ps(k)\window^2(k)}
  \\ \mbox{} \times
  \sum_{\ell_1, m_1, n_1} \sum_{\ell_2, m_2, n_2} \R^{m_1}_{\ell_1|n_1}
  \R^{m_2\dag}_{\ell_2|n_2} Y_{\ell_1 m_1}(\theta,\phi)
  Y^\dag_{\ell_2,m_2}(\theta,\phi) \psi_{n_1}(k) \psi_{n_2}(k)
  \Bigg),
\end{eqnarray}

\noindent where the normalisation factor has been left aside and
can be recovered by demanding the integral over all values to be
equal to 1.

   The harmonics $Y_{\ell m}$ and $\psi_n$ integrate out of this
expression entirely, using the orthonormality relation
\eqref{harmonic:orthonormal} and the spherical harmonic
completeness relation \eqref{harmonic:completness}.
Moreover, after rewriting the $a$ and $b$ coefficients with $m<0$
in terms of the $m>0$ coefficients, we obtain
\begin{equation}
  \G[\R] = \exp\Bigg( - \frac{1}{2\pi^2 (2\pi)^3}
  \sum_{\ell =0}^\infty \sum_{m=1}^{\ell} \sum_{n=1}^\infty
  |a^m_{\ell|n}|^2 + |b^m_{\ell|n}|^2 -
  \frac{1}{4\pi^2 (2\pi)^3} \sum_{\substack{\ell = 0 \cr \even{\ell}}}
  ^\infty \sum_{n=1}^\infty |a^0_{\ell|n}|^2 + |b^0_{\ell+1|n}|^2
  \Bigg) . \label{gaussian:coefficients}
\end{equation}

\noindent which is the gaussian expression presented in \eqref{2.2}.

In order to find the probability for given values of the central
amplitude $\R(r =0) = \vartheta_0$, we integrate
$\G[\R]$ with the $\diracd$-function factor in \eqref{2.3}. We
introduce the Fourier representation of the $\diracd$-function to write,
\begin{equation}
  \Prob(\vartheta_0) \propto \int [\d \R] \int_{-\infty}^\infty
  \d z \; \G[\R] \exp \left[ \imag z \left( \sum_{n=1}^\infty a^0_{0|n}
  \Sigma_n - \frac{(2\pi)^3}{\sqrt{4\pi}} \vartheta_0 \right) \right] ,
\end{equation}

\noindent where the functional measure is understood to be Eq.~\eqref{2.1}.
The final answer is obtained by integrating out $z$ together with
all of the $a$ and $b$ coefficients. In order to achieve this, it
is necessary to decouple $a^0_{0|n}$, $z$ and $\vartheta_0$ from each
other by successively completing the square in $a^0_{0|0}$ and
$z$. Working with $a^0_{0|0}$ first, we find
\begin{align}
  \nonumber
  &\exp\left( - \frac{1}{4\pi^2}  \frac{1}{(2\pi)^3} \sum_{n=1}^\infty
  |a^0_{0|n}|^2 + \imag z \sum_{n=1}^\infty a^0_{0|n} \Sigma_n \right) \\
  \label{gauss:asquare}
  = & \exp \left[ - \frac{1}{4\pi^2} \frac{1}{(2\pi)^3} \sum_{n=1}^\infty
  \left(a^0_{0|n} -  \imag 2 \pi^2 (2\pi)^3 z \Sigma_n \right)^2 -
  (2\pi)^3 \pi^2 z^2 \Sigma_{(2)}^2 \right] ,
\end{align}

\noindent where we have introduced a function $\Sigma_{(2)}^2$, defined by
$\Sigma_{(2)}^2 = \sum _{n=1}^\infty \Sigma_n^2$. In the final
probability distribution, $\Sigma_{(2)}^2$ will turn out to be the
variance of $\R(0)$. From Eq.~\eqref{gauss:asquare}, it is clear
that making the transformation $a^0_{0|n} \mapsto a^0_{0|n} +
 \imag 2\pi^2 (2\pi)^3 z \Sigma_n$ suffices to decouple $a^0_{0|n}$
from $z$. The measure, Eq.~\eqref{2.1}, is formally invariant
under this transformation. We can also complete squares for the
variables $z$ and $\vartheta_0$, giving 
\begin{equation}
  \exp \left( - (2\pi)^3 \pi^2 z^2 \Sigma_{(2)}^2 -
   \imag \frac{(2\pi)^3}{\sqrt{4\pi}}\vartheta_0 z \right) = \exp \left[ -
  (2\pi)^3 \pi^2 \Sigma_{(2)}^2 \left( z + \imag \frac{ \vartheta_0}{2 \pi^2
  \sqrt{4\pi} \Sigma_{(2)}^2} \right)^2 - \frac{\vartheta_0^2}{2\Sigma^2}
  \right] . 
\end{equation}

\noindent As before, the finite shift $z \mapsto z - \imag \vartheta_0 / 2\pi^2
\sqrt{4\pi} \Sigma_{(2)}^2$ leaves the measure intact and decouples $z$
and $\vartheta_0$. The $a$, $b$ and $z$ integrals can be done
independently, but since they do not involve $\vartheta_0$ they
contribute only an irrelevant normalisation to $\Prob(\vartheta_0)$.
The result is the Gaussian distribution for $\vartheta_0$,
\begin{equation}
  \label{gauss:gauss}
  \Prob(\vartheta_0) \propto \exp \left( -
  \frac{\vartheta_0^2}{2\Sigma_{(2)}^2} \right) . 
\end{equation}

\noindent It remains to evaluate the variance $\Sigma_{(2)}^2$. In the present
case, we have $\Sigma_n = \int_0^\Lambda \d k \, k^2 \psi_n(k)$.
From the completeness relation Eq.~\eqref{harmonic:complete}, it
follows that
\begin{equation}
  \sum_n k_0^2 \psi_n(k_0) k^2 \psi_n(k) = \frac{k^2 \ps(k_0)
  \window^2(k_0)} {k_0^3} \diracd(k-k_0) .\label{psi:orthogonality}
\end{equation}

\noindent $\Sigma_{(2)}^2$ is now obtained by integrating term-by-term
under the summation. The result coincides with the conventional
{smoothed} variance, 
\begin{equation}
  \label{total:variance}
  \Sigma^2_{\Lambda}(\kmax)
  = \int_0^\Lambda \d \ln k \; \window^2(k;\kmax) \ps(k) .
\end{equation}

\noindent Thus, as expected, Eq.~\eqref{gauss:gauss} reproduces the Gaussian
distribution \eqref{gaussian:prob} which was derived on the basis
of the central limit theorem, with the proviso that the parameters
(such as $\Sigma_{(2)}^2$) describing the distribution of $\vartheta_0$ are
associated with the smoothed field $\R$.

For the case of the central second derivative we integrate this
probability against the $\delta$-function containing the desired
condition \eqref{1.6} 
   \begin{equation}
     \mathbb{P}(\vartheta_2) = \int[d\R]
     \G[\R] \delta\left[\R''(0) - \vartheta_2
     \right]. \label{2:2}
   \end{equation}

\noindent Using again the expression of the $\delta$-function as an
 integral and the condition on the derivative in terms of the
 spherical harmonic coefficients we have:
 \begin{align}
   \mathbb{P}(\vartheta_2) \propto \int \,[d \R]\int \,dz
   \,\G[\R] \exp\left[\mathrm{i}z \left(
   \frac{3(2\pi)^3\vartheta_2}{\sqrt{4\pi}} 
   +\sum_n  \Sigma_{n}^{(4)}\left(\sqrt{\frac{4}{5}} a^0_{2|n} + a^0_{0|n}\right)
   \right)\right], \label{proba:ddr}
 \end{align}

\noindent where the factor $\Sigma_n^{(4)}$ is defined as
\begin{align}
  \Sigma_n^{(4)} = \int \,dk \,k^4 \psi_n(k).  \label{sigma:four}
\end{align}

\noindent   In the integral \eqref{proba:ddr} the terms with factors
of $a^0_{0|n}$ are 
   \begin{align}
     \exp\left[- \frac{1}{4\pi^2(2\pi)^3}\sum_{n=1}^\infty |a^0_{0|n}| +
     \mathrm{i}z \sum_{n=1}^\infty |a^0_{0|n}|\Sigma_n^{(4)}\right].
   \end{align}

  \noindent Completing squares, this last expression is equal to
  \begin{align}
  \exp\left[- \frac{1}{4\pi^2(2\pi)^3}\sum_{n=1}^\infty \left(|a^0_{0|n}| -
     \mathrm{i}(2\pi)^3 2\pi^2 z \Sigma_n^{(4)}\right)^2 - (2\pi)^3
     \pi^2 z^2 \Sigma_{(4)}^2\right]. \label{compsq:a0}    
  \end{align}

\noindent  In the same way we can complete squares for the expansion factors
  $a^0_{2|n}$:
\begin{align}
     &\exp\left[-\frac{1}{4\pi^2(2\pi)^3}\sum_{n=1}^\infty |a^0_{2|n}| +
     \mathrm{i}z \frac{4}{5}\sum_{n=1}^\infty
     |a^0_{2|n}|\Sigma_n^{(4)}\right]\notag\\
     &\qquad = \exp\left[-\frac{1}{4\pi^2(2\pi)^3}\sum_{n=1}^\infty
     \left(|a^0_{0|n}| - \imag (2\pi)^3 \frac{4\pi^2}{\sqrt{ 5}}z
     \Sigma_n^{(4)}\right)^2 - (2\pi)^3 \pi^2 \frac{4}{5} z^2
     \Sigma_{(4)}^2\right]. \label{compsq:a2}
   \end{align}

\noindent  And finally one can also complete squares for the terms containing
the variable $z$ which are independent of $a^0_{0|n}$ and $a^0_{2|n}$,
\begin{align}
  \exp&\left[ -
    (2\pi)^3\pi^2\left(\frac{9}{5}\right)z^2\Sigma_{(4)}^{2}  +
    \mathrm{i}\frac{3 (2\pi)^3}{\sqrt{4\pi}}  \vartheta_2 z \right]  =
    \notag\\ 
    \exp& \left[- (2\pi)^3 \pi^2
    \left(\frac{9}{5}\right)\Sigma_{(4)}^{2} \left(z - 
    \mathrm{i}\frac{5 }{12\sqrt{\pi^5}}
    \frac{\vartheta_2}{\Sigma_{(4)}^2} \right)^2 - \frac{5}{2}
    \frac{\vartheta_2}{\Sigma_{(4)}^{2}}\right],
\end{align}

\noindent where for simplification we have written
\begin{align}
  \Sigma_{(4)}^2 =  \sum_{n=1}^{\infty}\left(
  \Sigma_n^{(4)}\right)^2.   \label{sumsigma:four} 
\end{align}

\noindent To evaluate this variance of the second derivative  we use
again the property \eqref{psi:orthogonality} to integrate the complete
summation and obtain 
\begin{align}
  \Sigma_{(4)}^2 = \int_0^{\Lambda} d\ln{k} \window^2(k,k_H)\ps(k)\,k^4.
\end{align}
\noindent  So by making the change of variables
\begin{align}
  a^0_{0|n} \mapsto & a^0_{0|n}  + \mathrm{i} 2   \pi^2 (2\pi)^3
  \Sigma_n^{(4)} z,\\ 
  a^0_{2|n} \mapsto & a^0_{2|n}  +  \mathrm{i} \frac{4  \pi^2}{\sqrt{5}} 
  (2\pi)^3 \Sigma_n^{(4)} z\\
  \mbox{and } \qquad z \mapsto & z + \mathrm{i}\frac{5}{12 \sqrt{\pi^5}}
  \frac{\vartheta_2}{\Sigma_{(4)}^{2}},
\end{align}
we can perform all the integrals and eliminate the gaussian ones
which contribute only up to an overall numerical factor subsequently
  absorbed by normalisation. The remaining factor expresses the
  probability of finding a perturbation $\R$ with a central second
  derivative of value $\vartheta_2$,

  \begin{align}
    \mathbb{P}\left[ \R''(\br = 0 ) = \vartheta_2 \right] \propto \exp\left(-
    \frac{5 \vartheta_2^2}{2 \Sigma_{(4)}^{2}}\right).
  \end{align}

  \end{document}